\documentclass[twoside]{article}
\usepackage{a4wide}
\usepackage{amsmath}
\usepackage{amsfonts}
\usepackage{amssymb}
\usepackage{times}
\def\date{October, 2004}
\newcommand{\ed}{\end{document}}
\newcounter{mycnt}[section]

\def\!{\kern -0.15ex}

\begin{document}
\title{Resolution of the $GL(3) \supset O(3)$ state labelling problem
via $O(3)$-invariant Bethe subalgebra of the twisted Yangian}
\author{P D Jarvis and R B Zhang}
{ {\renewcommand{\thefootnote}{\fnsymbol{footnote}}
\footnotetext{\kern-15.3pt AMS Subject Classification: 17B10,
17B35, 22E47.} }} \maketitle

\begin{abstract}
The labelling of states of irreducible representations of $GL(3)$
in an $O(3)$ basis is well known to require the addition of a
single $O(3)$-invariant operator, to the standard diagonalisable
set of Casimir operators in the subgroup chain $GL(3) \supset
O(3)\supset O(2)$.  Moreover, this `missing label' operator must
be a function of the two independent cubic and quartic invariants
which can be constructed in terms of the angular momentum vector
and the quadrupole tensor.  It is pointed out that there is a
unique (in a well-defined sense) combination of these which
belongs to the $O(3)$ invariant Bethe subalgebra of the twisted
Yangian $Y(GL(3);O(3))$ in the enveloping algebra of $GL(3)$.

\end{abstract}
\pagebreak

The necessity for using adapted bases compatible with kinematical
or dynamical symmetries of a quantum system has long been
recognised as an essential tool for dealing with the implications
of the symmetries in terms of constraints on physical quantities,
and for reducing the number of computations.  Unfortunately the
symmetries of interest in physics are not always those which admit
`canonical' bases, possessing complete sets of orthonormal basis
functions which are common eigenfunctions of a maximal set of
commuting Casimir operators,  in a chain of subgroups extending
from the group of interest up to the maximal (unitary) group of
transformations on the space in question. A case of a
serendipitous group labelling occurred in Racah's early work on
equivalent $f$ electrons for rare earth spectra \cite{Racah1964}
where the exceptional group $G_{2}$ was found to extend the
symmetry adapted group-subgroup chain from $U(7) \supset O(7)
\supset O(3)$ to $U(7) \supset O(7) \supset G_{2} \supset O(3)$,
in a way which made the specifications of the electronic wave
functions complete for the cases studied.  In fact all steps in
this chain, and generically $U(N) \supset O(N)$, are of the type
where there are one or more `missing labels' -- in order to
resolve multiplicities in the restriction of irreducible
representations of the larger group to the subgroup, the
degeneracies should be removed with the help of additional
subgroup invariant operators which cannot be Casimir operators of
the group or the subgroup, but must be invariant operators taken
in the enveloping algebra of the group that commute with each
other. It follows from general arguments based on the double
commutant theorem that there always exist enough such invariant
operators to provide the missing labels for resolving
multiplicities.

In this note we study the case of $N=3$, namely the classic `$U(3)
\supset O(3)$ state labelling problem', which is ubiquitous in
atomic, nuclear and many body physics.  Below, we give a brief
introduction to the notation necessary to define the problem, and
we review the known result that there are two admissible
additional $O(3)$ invariant operators in the enveloping algebra of
$U(3)$ which are candidates for the single `missing label' needed
in this case.  We then turn to the formalism of Yangian algebras,
which provide a powerful way of handling the (infinite
dimensional) enveloping algebras of the classical Lie algebras.
Specifically we consider the so-called Bethe subalgebras
\cite{Molev2002, NazarovOlshanski1995}, which are maximal
commutative subalgebras. For the $N=3$ case we show explicitly
that there is a \textit{unique} (in a well-defined sense)
combination of the candidate $O(3)$ non-subgroup invariants, which
belongs to the $O(3)$ invariant Bethe subalgebra of the twisted
Yangian $Y(GL(3);O(3))$.

\bigskip

The $SU(3) \supset O(3)$ state labelling problem is comprehensively
examined in \cite{JuddMillerPateraWinternitz1974}, which also includes
extensive numerical evaluations for low dimensional representations.
In the following we work with the complex algebras, and so refer to
$GL(3)$, $O(3)$ and so on (as well as following the physics convention
of not distinguishing notationally between group and algebra).

Consider the standard generators $E_{ij}, i,j = 1,2,3$ of $GL(3)$
with commutation relations modelled on those of the defining $3
\times 3$ matrix units $e_{ij}$, acting on 3 basis vectors $e_{i}$
in the usual way:
\begin{align}
    \label{eq:GL3CommutationRelations}
    {[}E_{ij}, E_{kl}{]} = &\,  \delta_{jk}E_{il} - \delta_{il} E_{kj}.
\end{align}
For applications in many body physics the orbital angular momentum
generators are given by
\begin{align}
    \label{eq:O(N)Definition}
    L_{ij} = & \,E_{ij} - E_{ji}
\end{align}
from which the usual vector angular momentum generators follow as
$L_{1}=L_{23}$, $L_{2}=L_{31}$, $L_{3} = L_{12}$. For labelling
states in irreducible representations of $GL(3)$ one requires a
maximal set of commuting operators. Those in the group-subgroup
chain $GL(3) \supset O(3) \supset O(2)$ are the associated Casimir
operators. Taking them to be the standard Gel'fand invariants
gives for $GL(3)$, at increasing degree,
\begin{align}
    \label{eq:GL3Casimirs}
    C^{(1)}= & \, \sum_{i=1}^{3} E_{ii}, \nonumber \\
    C^{(2)} =&\, \sum _{i,j=1}^{3} E_{ij}E_{ji}, \quad \mbox{and}
    \nonumber \\
    C^{(3)} =& \,\sum _{i,j,k=1}^{3} E_{ij}E_{jk}E_{ki},
\end{align}
the linear Casimir being of course the number operator
$N \equiv  E_{11}+E_{22}+E_{33}$ (which determines the energy $(N+\frac
32)\hbar \omega$ if the system is a three dimensional isotropic
oscillator). For ease of writing we adopt the notation $\langle E
\rangle$, $\langle E^{2} \rangle$ and $\langle E^{3} \rangle$ for
these Casimir operators. Further invariants are the quadratic Casimir
for $O(3)$,
\begin{align}
    C^{[2]} = &\,\sum_{i,j =1}^{3} L_{ij}L_{ji},
\end{align}
which we denote by $\langle L^{2}\rangle$, and
the $O(2)$ angular momentum component $L_{12}$ (the Casimir operator
being $|L_{12}|$).

The remaining
$GL(3)$ generators are the quadrupole tensor
\begin{align}
    Q_{ij} = &\, E_{ij} + E_{ji}, \quad \mbox{or} \quad {Q'}_{ij}  = E_{ij} +
    E_{ji} - \frac 23 N \delta_{ij},
\end{align}
where the traceless form $Q' \in SL(3)$ separates the number operator $N$.
It is proven in \cite{JuddMillerPateraWinternitz1974} using
methods of invariant theory (or so-called `integrity bases')
that the algebraically independent
$O(3)$ invariants in the $GL(3)$ enveloping algebra are at degree 3,
4 and 6 and can be taken to be
\begin{align}
    \label{eq:X346Defn}
    X^{(3)} = &\, \langle LQ'L \rangle =
    \sum_{i,j,k =1}^{3} L_{ij}{Q'}_{jk}L_{ki}, \qquad
    X^{(4)} = \langle L{Q'}^{2}L\rangle =
    \sum_{i,j,k,l =1}^{3} L_{ij}{Q'}_{jk}{Q'}_{kl}L_{li}, \quad
    \mbox{and} \nonumber \\
    X^{(6)} = & \,\sum_{i,j,k,l,m,n =1}^{3}
    L_{ij}L_{kl}L_{mn}{Q'}_{ik}{Q'}_{lm}{Q'}_{jn}.
\end{align}
In fact $X^{(3)}$ and $X^{(4)}$ are primary invariants, while
$X^{(6)}$ is secondary in that $(X^{(6)})^2$ is a polynomial in
$X^{(3)}$, $X^{(4)}$ and terms of lower degree (the commutator
${[} X^{(3)},X^{(4)}{]}$ gives essentially $X^{(6)}$ up to
invariants of lower degree). Let $Z(GL(3); O(3))$ be the
commutative subalgebra of $U(GL(3))$ generated by the Casimir
operators of $GL(3)$ and $ O(3)$, namely, the operators $\langle E
\rangle$, $\langle E^2 \rangle$, $\langle E^3 \rangle$, and
$\langle L^2 \rangle$.  Then \textit{any} $O(3)$ invariant
operator that can be used for state labelling must be of the form
$f(X^{(3)},X^{(4)}) + X^{(6)}g(X^{(3)},X^{(4)})$ for some
polynomials $f$, $g$ with coefficients in  $Z(GL(3); O(3))$. As
the object of interest is the restriction of an irreducible
$GL(3)$ representation to $O(3)$, the $GL(3)$ Casimir operators,
all being scalar multiples of the identity, lend no help for
resolving multiplicities of $O(3)$ irreducible representations. It
is the $O(3)$ Casimir and invariant operators like $ X^{(3)}$ and
$X^{(4)}$ that provide the desired information for state
labelling. In \cite{JuddMillerPateraWinternitz1974}, explicit
numerical evaluations of $X^{(3)}$ and $X^{(4)}$ are tabulated for
irreducible representations of $SU(3)$ of high enough dimensions
that multiplicities up to 3 occur in the restriction to $SO(3)$.

\bigskip

The infinite dimensional Yangian algebras have been intensively
studied in relation with applications to integrable systems and
the inverse scattering method.  They have a very remarkable
formulation as noncommutative matrices over the ring of formal
Laurent series, enabling the combinatorics of the coefficients
involved in commutation relations and other constructs such as
invariants and coproducts, to be handled by appropriate shifts of
the formal variable $u$.  A significant identification is that of
the generators of the Yangian $Y(GL(N))$ with elements of the
enveloping algebra of $GL(N)$. Denoting the generators of the
Yangian by $ {t^{(m)}}_{ij}$, $m= 0,1,2,\ldots$, this entails
\begin{align}
    {T}_{ij}(u) =& \,\sum_{m=0}^{\infty} (-1)^{m} \frac{   {t^{(m)}}_{ij}   }{u^{m}},
    \quad { t^{(m)} }_{ij} = (E^{m})_{ij}, \quad \mbox{where} \nonumber \\
    (E^{0})_{ij} = & \,\delta_{ij}, \quad \mbox{and} \quad
    (E^{m})_{ij} = \sum_{k=1}^{N}E_{ik}(E^{m-1})_{kj} \quad
    \mbox{for} \quad m>0.
\end{align}
More simply, the \textit{inverse} of this series provides the evaluation homomorphism
\begin{align}
    T_{ij}(u) \rightarrow & \,\delta_{ij} + \frac{E_{ij}}{u}.
\end{align}
from $Y(GL(N))$ to the universal enveloping algebra $U(GL(N))$.
The generators are succinctly
written with the Laurent series in matrix form,
\begin{align}
    \label{eq:TDefn}
    T(u) = &\, \sum_{i,j=1}^{N} e_{ij}\otimes T_{ij}(u)
\end{align}
regarded as an
$N \times N$ matrix with entries in the Yangian, or an element of
$End({\mathbb C}^{N}) \otimes Y(GL(N))[[u]] $.
Central to such manipulations is the $R$-matrix (an operator on
${\mathbb C}^{N}\otimes {\mathbb C}^{N}$),
\begin{align}
    R(u) = &\, u\cdot 1 + {\mathbb P},
\end{align}
where ${\mathbb P}$ is the permutation operator defined by ${\mathbb P}
\, e_{i} \otimes e_{j} = e_{j} \otimes e_{i}$.
In terms of elementary matrices
\begin{align}
    {\mathbb P} = &\,\sum_{i,j=1}^{N} e_{ij} \otimes e_{ji}.
\end{align}

The twisted Yangian which we denote $Y(GL(N);O(N))$ with generators $S(u)_{ij} \in
Y(GL(N))[[u]]$ associated
with the above embedding (\ref{eq:O(N)Definition}) of $O(N)$
in $GL(N)$ is introduced as
\begin{align}
    S_{ij}(u) = & \,\sum_{m=0}^{\infty} \frac{   {s^{(m)}}_{ij} }{u^{m}}, \nonumber \\
    S(u) = &\, \sum_{i,j=1}^{N} e_{ij}\otimes S_{ij}(u), \nonumber \\
    S(u) = & \,T(u)\widetilde{T}(-u) = 1 +\frac{E - \widetilde{E}}{u} -
    \frac{E \widetilde{E}}{u^{2}},
\end{align}
using the definition (\ref{eq:TDefn}) above for $T(u)$, and with
$\widetilde{T}_{ij}(u) = {T}_{ji}(u)$. Here $E=\sum e_{i j}\otimes
E_{i j}$, and $\tilde E=\sum e_{i j}\otimes E_{j i}$.  The
relevant $R$-matrix is now the partial transpose
\begin{align}
    \widetilde{R}(u) = &\, u \cdot 1 + {\mathbb Q},
\end{align}
where ${\mathbb Q}$ is the projection operator onto the
$1$-dimensional $O(N)$-submodule ${\mathbb C}\sum_i e_i\otimes
e_i.$ In terms of elementary matrices
\begin{align}
    {\mathbb Q} = &\,\sum_{i,j=1}^{N} e_{ij} \otimes e_{ij}.
\end{align}
As with the
$GL(N)$ Yangian, the commutation relations can be succinctly expressed
using the $\widetilde{R}$-matrix, and many structural properties of the
algebra established (see \cite{Molev2002,NazarovOlshanski1995}).

One of the most fundamental aspects of the Yangian is the fact that
the \textit{trace} $tr[T(u)]$ of the Laurent series over $End({\mathbb C}^{N})$,
namely
\begin{align}
    tr\left[ T(u) \right] = &\, \sum_{i=1}^{N} T_{ii}(u),
\end{align}
commutes with $tr[T(v)]$ for arbitrary $u,v$ -- that is, the
coefficients provide an infinite set of commuting operators.  The
diagonalisability of the transfer matrix is of course the underpinning
of many of the applications of Yangians and the Yang-Baxter equation
to integrable systems.  The same property can also be proved for the
trace of the twisted Yangian, $tr[S(u)]$.

The identification of abelian subalgebras is not limited solely to the
trace of the Yangian however.  The so-called `quantum determinant' is an
object whose index structure is that of a determinant (of the Yangian
matrix), but whose terms involve systematic shifts in the formal
variable in order to compensate for the non-commutativity.  The
quantum determinant is thus a Laurent series in principle encoding an infinite
number of coefficients, which this time belong to the \textit{centre}
of the algebra. Beyond the trace and the quantum determinant there is
a remarkable set of infinite dimensional abelian subalgebras indexed
by a fixed $N \times N$ matrix $Z=\sum_{i,j}e_{ij}z_{ij}$, the
so-called Bethe subalgebras \cite{Molev2002,NazarovOlshanski1995}
$B(GL(N);Z)$ and $B(O(N);Z)$. If $Z$ is
generic, in that it has a simple spectrum, these subalgebras are maximal,
and generate the equivalent for the Yangians, of a complete set of
commuting labelling operators for representations of finite-dimensional
simple Lie algebras.

Henceforth we specialize to the $GL(3) \supset O(3)$ case, and
give the concrete constructions for the twisted Yangian
$Y(GL(3);O(3))$. Because we are interested only in $O(3)$
invariants, we consider $Z=1_{3\times 3}$ and the Bethe subalgebra
$B(O(3); 1_{3 \times 3})$. This choice of $Z$ simplifies the
general definitions, and the generators of the corresponding Bethe
subalgebra are the coefficients in $u^{-1}$, $u^{-2}$, $\cdots$ of
the following three elements $A_{1}(u)$, $A_{2}(u)$, $A_{3}(u)$:
\begin{align}
    A_{1}(u) =&\,tr_{1}\left[ S_{1}(u-1) \right], \nonumber \\
    A_{2}(u) =& \,tr_{12} \left[ {\mathbb A}_{12} \otimes 1
    \cdot S_{1}(u-1) \widetilde{R}_{12}(-2u+3)
                               S_{2}(u-2) \right], \nonumber \\
    A_{3}(u) =& \,tr_{123} \left[ {\mathbb A}_{123} \otimes 1
    \cdot S_{1}(u-1) \widetilde{R}_{12}(-2u+3)
                           \widetilde{R}_{13}(-2u+4)S_{2}(u-2)
                           \widetilde{R}_{23}(-2u+5) S_{3}(u-3)\right].
\end{align}
The subscript notation indicates to which of the subspaces various
objects belong. For example in $A_{2}$, $S_{2}(u) = \sum_{i,j}1
\otimes e_{ij} \otimes S_{ij}(u)$, whereas in $A_{3}$, $S_{2}(u) =
\sum_{i,j}1 \otimes e_{ij} \otimes 1 \otimes  S_{ij}(u)$. The
${\mathbb A}$ are antisymmetrisation operators acting on the
appropriate spaces, with
\begin{align}
   {\mathbb A}_{12} =& \,1 - {\mathbb P}_{12}, \nonumber \\
   {\mathbb A}_{123} =& \,1 - {\mathbb P}_{12} - {\mathbb P}_{13} - {\mathbb
   P}_{23} + {\mathbb P}_{12}{\mathbb P}_{23} + {\mathbb P}_{13}{\mathbb
   P}_{23}.
   \label{eq:A123Defn}
\end{align}
Finally, the matrix objects are subjected to a total trace.

In $A_{1}(u)$ we recognise the basic transfer matrix trace
discussed already, merely rewritten to emphasise its relationship
to its partners in the Bethe subalgebra.  Also, the top member
$A_{3}(u)$ is the quantum determinant itself (always present, and
independent of the matrix $Z$, because it is associated with the
centre).  In fact for the twisted Yangian, the quantum determinant
is essentially the square of the quantum determinant for the
Yangian itself.  For the present $N=3$ case we can thus compute
$A_{3}(u)$ via
\begin{align}
    A_{3}(u) = & \,B_{3}(u)B_{3}(4-u), \qquad \mbox{where} \nonumber \\
    B_{3}(u) = & \,tr_{123} \left[ {\mathbb
    A}_{123} \otimes 1 \cdot T_{1}(u-1)T_{2}(u-2)T_{3}(u-3) \right].
    \label{eq:B3Defn}
\end{align}

>From (\ref{eq:A123Defn}), (\ref{eq:B3Defn}) and the previous
definitions it is straightforward to compute these Bethe subalgebra
generators in terms of traces of polynomials in the $GL(3)$ and $O(3)$
generators as in (\ref{eq:GL3Casimirs}) above.  We find explicitly
\begin{align}
    \label{eq:A123Eval}
    A_{1}(u) = &\, 3 - \frac{\langle E \widetilde{E} \rangle }{(u-1)^{2}} ,
    \nonumber \\
    A_{2}'(u) =&\langle L ^2\rangle + \frac{\langle E \tilde E E
\tilde E \rangle-\langle L ^2\rangle}{(u-1)(u-2)},
    \nonumber \\
    B_{3}(u) =&\, 6  - 2{\langle E
    \rangle}\left(\frac{1}{(u-1)} +\frac{1}{(u-2)}+\frac{1}{(u-3)} \right)\nonumber \\
    &-3\frac{\langle E^{2}\rangle - \langle E \rangle^{2}}
    {(u-1)(u-3)}
    - \frac{2 \langle E^{3}\rangle + (\langle E \rangle +1) (\langle E \rangle^2 - 3
     \langle E^{2} \rangle)}{(u-1)(u-2)(u-3)}.
\end{align}
where we have defined the essential part of $A_{2}(u)$ after
making combinations with $A_1(u)$ by
\begin{align}
    A_{2}'(u)= (u-1)(u-2)\left(\frac{A_2(u)}{2u-3} +A_1(u)
A_1(u-1) - A_1(u) - A_1(u-1) +3\right).
\end{align}
In simplifying the expressions basic symmetry properties have been
used, for example $L_{ij}=-L_{ji}$ giving $\langle L \rangle =0$,
and from the definition of the angular momentum operators $\langle
L^{2} \rangle = 2\langle E^{2} \rangle - 2 \langle E \widetilde{E}
\rangle$. Similarly
\begin{align}
    {(E \widetilde{E})}_{ij} - {(E \widetilde{E})}_{ji} = \, L_{ij}
\end{align}
upon using the commutation relations (\ref{eq:GL3CommutationRelations}), so that
\begin{align}
\langle E \widetilde{E} L \rangle = & \,\langle L E \widetilde{E}
\rangle = \textstyle{\frac 12} \langle L^{2} \rangle.
\end{align}
By taking appropriate linear combinations, the
independent generators of
$B(O(3);1_{3\times 3})$ can be taken to be the set
\begin{align}
\langle E \rangle^{2}, \langle E^{2} \rangle, \langle E \rangle\langle
E^{3} \rangle, \langle L^{2} \rangle, \quad \mbox{and } \quad
\langle E \widetilde{E} E \widetilde{E} \rangle.
\end{align}
A useful way to look at the Bethe subalgebra $B(O(3);1_{3\times
3})$ is to consider it as an associative algebra over
$B(O(3);1_{3\times 3})\cap Z(GL(3); O(3))$ generated by the
operator $\langle E \widetilde{E} E \widetilde{E} \rangle.$

\bigskip

To complete the identification with labelling operators for $GL(3)
\supset O(3)$, it is necessary to re-write the operators
(\ref{eq:X346Defn}) in the trace notation as above.  In such
expressions, use must again be made of the commutation relations
in order to group similar terms.  Within strings of the form
$\langle X \rangle = \langle E^{m} \widetilde{E}^{n} \cdots
\rangle$ for example, simplifications that can be made are that
the trace is cyclic in nature, and also that the transpose
$\widetilde{X}$ behaves (anti)-involutively (of course, $\langle
\widetilde{X} \rangle = \langle {X} \rangle$) -- in both cases up
to rearrangements in the order of terms, which produce invariants
of lower degree after applying the commutation relations.  It
should also be noted that for $O(3)$,  $\langle L^3 \rangle =
\frac {1}{2} \langle L^2 \rangle$ and $\langle L^4 \rangle =
\frac{1}{2} \langle L^2 \rangle(\langle L^2 \rangle+2)$. Similarly
in $GL(3)$ the quartic Casimir $\langle E^{4} \rangle=tr(E^4)$ is
not algebraically independent, being of higher degree than the
exponents for invariants of the group, namely 1,2 and 3 for
$GL(3)$.  In this case by invoking the characteristic identity
\cite{Green1971,BrackenGreen1971} for the matrix $E_{ij}$ (the
analogue of the matrix Cayley-Hamilton identity, but with
coefficients in the centre of the enveloping algebra), one can
show that $\langle E^{4}\rangle$ is a linear combination of
$\langle E \rangle \langle E^{3} \rangle$ and similar reducible
terms of degree up to 4, consisting of products of traces with
lower degree.

Because of the discussion following equation (\ref{eq:X346Defn}),
we may consider, instead of $X^{(3)}$ and $X^{(4)}$ themselves,
their combinations over $Z(GL(3); O(3))$ defined by
\begin{align}
    Y^{(3)}= &X^{(3)}+ \frac{2}{3} C^{(1)} C^{[2]} ,\nonumber\\
    Y^{(4)} =&X^{(4)}+ \frac{4}{3} C^{(1)} X^{(3)} + \frac{4}{9}
    (C^{(1)})^2 C^{[2]}. \nonumber
\end{align}
Some very lengthy calculations yield
\begin{align}
    Y^{(3)} =&
    - (\langle E \widetilde{E}^2 \rangle + \langle \widetilde{E} E^2 \rangle)
    +2\langle E^3 \rangle -3\langle E^2 \rangle +
    \langle E\rangle^2, \nonumber \\
    Y^{(4)} =&-2 \langle E \widetilde{E} E \widetilde{E}\rangle +
     2 \langle E^4 \rangle -6 \langle E^3 \rangle + 2\langle E \rangle\langle E^2 \rangle
    + 6 \langle L^2 \rangle. \nonumber
\end{align}
Clearly the two algebraically independent invariants equivalent to
$X^{(3)}$ and $X^{(4)}$ in the trace notation are $\langle E
\widetilde{E}^2 \rangle + \langle \widetilde{E} E^2 \rangle $, and
$\langle E \widetilde{E} E \widetilde{E} \rangle$. Moreover, the
cubic invariant piece $\langle E \widetilde{E}^2 \rangle + \langle
\widetilde{E} E^2 \rangle $, which does not belong to $B(O(3);
1_{3 \times 3})$, is completely eliminated from $Y^{(4)}$. We
further define
\begin{align}
 Y=&X^{(4)}+ \frac{4}{3} C^{(1)} X^{(3)} + \frac{4}{9}
    (C^{(1)})^2 C^{[2]} - \left(2 \langle E^4 \rangle -
    6 \langle E^3 \rangle + 2\langle E \rangle\langle E^2 \rangle
    + 6 \langle L^2 \rangle \right). \label{eq:PreferredMissingLabel}
\end{align}
Then $$Y=-2\langle E \widetilde{E} E \widetilde{E}\rangle.$$ This
preferred labelling operator is the \textit{unique} (up to
elements of $B(O(3); 1_{3 \times 3})\cap Z(GL(3); O(3))$ and
complex scalar multiples) linear combination over $Z(GL(3); O(3))$
of the invariants of \cite{JuddMillerPateraWinternitz1974} at
order $4$, which belongs to the $O(3)$ invariant Bethe subalgebra
$B(O(3); 1_{3 \times 3})$.

\bigskip

In this note we have pointed out that the `missing label' in the
$GL(3)\supset O(3)$ group reduction can be identified with the
appropriate generator of the $O(3)$ invariant Bethe subalgebra of
the twisted Yangian in the $GL(3)$ enveloping algebra.  This
identification answers the longstanding puzzle of the lack of any
systematic way to resolve the labelling problem, and casts light
on known results, such as Racah's proof (\cite{Racah1964}, cited
in \cite{JuddMillerPateraWinternitz1974}) that there is
\textit{no} choice of hermitean labelling operator with a rational
spectrum.  In general terms, it provides an interesting insight
into conventional group representation theory, coming ultimately
from the study of integrable systems in physics (see
\cite{Baxter1982}).

Our present result can clearly be generalised in several
directions. It is known for example that higher dimensional
analogues of the $N=3$ case have quadratically growing numbers of
`missing labels', for example 2 for $GL(4) \supset O(4)$, 4 for
$GL(5) \supset O(5)$ and so on \cite{Jarvis1974}.  It is tempting
to conjecture that also in these cases, the invariant Bethe
subalgebra of the twisted Yangian will provide a sufficient set of
commuting labelling operators.  Along these lines one can further
extend the analysis to other labelling problems, for example
$Sp(4) \supset Sp(2)$, or even to exceptional embeddings such as
that of $G_{2} \supset SO(3)$ mentioned in the introduction.  For
the $U(3) \supset O(3)$ labelling problem itself, there is of
course the task of numerical evaluation and analysis of the
spectrum of preferred missing label
(\ref{eq:PreferredMissingLabel}), in the light of the present
framework.  Further work along these lines is in progress.

\paragraph{Acknowledgement:} We would like to thank Dr Alexander Molev for
discussions on Yangians. This work is supported by grants
DP0208808, DP0211311, and DP0451790 from the Australian Research
Council.


\bigskip

\noindent
{\sc
Peter D Jarvis, University of Tasmania, School of Mathematics and Physics,
GPO Box 252C, 7001 Hobart, TAS, Australia,
{\small\tt Peter.Jarvis@utas.edu.au} \\
Ruibin Zhang, School of of Mathematics and Statistics, University of
Sydney, NSW 2006, Australia, {\small\tt rzhang@maths.usyd.edu.au} \\
Report Number UTAS-PHYS-2004- }
\end{document}